\documentclass[%%preprint-reprint, twocolumn,
 preprint,
 amsmath,amssymb,
 aps,
]{revtex4-1}

\usepackage{graphicx}% Include figure files
\usepackage{dcolumn}% Align table columns on decimal point
\usepackage{bm}% bold math
\begin{document}

%\preprint{APS/123-QED}

\title{Uncited papers are not unread\\}% Force line breaks with \\

\author{Michael Golosovsky}
\email{michael.golosovsky@mail.huji.ac.il}
%\author{Sorin Solomon}%
\affiliation{The Racah Institute of Physics, The Hebrew University of Jerusalem, 9190401 Jerusalem, Israel\\
}%
\date{\today}% It is always \today, today,
\begin{abstract}
We study citation dynamics of the Physics, Economics, and Mathematics papers published in 1984 and focus on the  fraction of uncited papers in these three collections.  Our  model of citation dynamics, which considers citation process as an inhomogeneous Poisson process, captures this   uncitedness ratio fairly well. It should be noted that all parameters and variables in our model  are related to citations and their dynamics, while uncited papers appear as a byproduct of the citation process and this is the Poisson statistics which makes the cited and uncited papers inseparable. This indicates that the most part of uncited papers constitute the inherent part of the scientific enterprise, namely, uncited papers are not unread.
\begin{description}
\item[PACS numbers]01.75.+m, 02.50.Ey, 89.75.Fb, 89.75.Hc
\end{description}
\end{abstract}
\pacs{{01.75.+m, 02.50.Ey, 89.75.Fb, 89.75.Hc}}% PACS, the Physics and Astronomy
\keywords{Suggested keywords}%Use showkeys class option if keyword
\maketitle

\section{Introduction}
The problem of uncited papers bothers the scientists since time immemorial. With the  appearance of the Science Citation Index  which allowed to count citations easily,  it became clear that about 10$\%$ of all research papers remain uncited \cite{SollaPrice1965}. This is  quite an appreciable fraction of the whole scientific enterprise and the natural question arose whether uncited papers are the burden to science or not, in other words, whether they are read or unread.  Seglen \cite{Seglen1992} realized  that the presence of uncited papers is an  inevitable consequence of statistics and discrete character of the citation process.  Therefore, the real question is whether the number of uncited papers is compatible with that expected from discrete statistics or not.

 The proper assessment of the phenomenon of uncitedness is important for scientific research policies.  Bibliometric scientists  contributed a large effort to empirical characterization of the number and composition of uncited papers, they studied how phenomenon of uncitedness  depends on the discipline, document kind, country, and year \cite{Wallace2009,Nicolaisen2019,Thelwall2016b,Leeuwen2005}. The measurements of uncitedness have been recently reviewed in \cite{Nicolaisen2019}  and  the thorough summary of the subject has been presented by van Noorden \cite{vanNoorden2017} and Lariviere and Sugimoto \cite{Sugimoto2018}.  It turns out that the uncitedness ratio, namely,  the fraction of uncited papers in a collection,  strongly depends on the observation time window, in such a way that  it is not even clear whether, for a given collection, the uncitedness ratio achieves some limiting value in the long time limit.   This indicates that the notion of uncitedness should be better understood and  characterized  before being used in the assessment of the productivity of science.

Existing theoretical models, which conceptualize the phenomenon of uncitedness, successfully  predict the uncitedness ratio in the collection of papers during first couple of years after publication  but fail to account for the uncitedness in the long time limit. In particular, van Leeuwen and Moed \cite{Leeuwen2005}  related uncitedness ratio for the journals to their impact factor (which  is nothing else but the mean number of citations per paper garnered in the first couple of years after publication); Hsu and Huang \cite{Hsu2012}, Burrell \cite{Burrell2013}, and Egghe \cite{Egghe2013} claimed a direct relation between the uncitedness ratio  and the mean number of citations for a collection of papers; while Wallace, Lariviere, and Gingras \cite{Wallace2009} demonstrated that uncitedness ratio is strongly affected by the fact that the number of publications and the average length of their reference lists both grow with time. Yet, a comprehensive study of Thelwall \cite{Thelwall2016b} showed relation between the uncitedness ratio and the shape of citation distribution.  Thus, while several factors affecting the uncitedness ratio were identified (mean number of citations, growth of the number publications and of the reference list length, the shape of the citation distribution), the existing models focused only on one or few  of these factors and on short observation time window comprising a couple of years after publication. The comprehensive model that includes all these factors and predicts the uncitedness ratio in the long time limit has been missing.

 We have recently developed a fully calibrated model of citation dynamics of research papers \cite{Golosovsky2017,Golosovsky2019}. Here, we apply this model to account for uncitedness ratio and to trace its relation to microscopic parameters which determine citation dynamics of papers.  We report here not only the model and but also the  measurements which were specially designed to verify our theoretical speculations. These measurements focus on  three disciplines: Physics, Economics, and Mathematics, and the publication year 1984. We demonstrate that the presence of uncited papers  is an inevitable consequence of discrete statistics, hence the uncited papers, at least  in these collections, are not useless.

\section{The model of citation dynamics and the  uncitedness ratio}
We present here a short summary of our model of citation dynamics \cite{Golosovsky2017,Golosovsky2019}  focusing on uncited  papers. Consider a paper $j$.    The  author publishing a new study may cite this paper after finding it in databases, in  scientific journals,  or following recommendations of colleagues or news portals. We name this a direct citation. An author of another new paper can find the paper $j$ in the reference list of one or several of his preselected  papers  and cite it as well. If the paper $j$ entered the reference list of a new paper as a result of such copying strategy, we name this indirect citation. Each direct citation of the paper $j$ triggers  cascades of indirect citations.

The model assumes that citation dynamics of a paper   follows an inhomogeneous Poisson process, namely, its  citation rate  in year $t$   has a probability distribution  $\frac{\lambda_{j}^{k_{j}}}{k_{j}!}e^{-\lambda_{j}}$ where $\lambda_{j}(t)$ is  the latent citation rate which is specific for each paper. It is a sum of the direct and indirect citation rates, namely, $\lambda_{j}(t)=\lambda_{j}^{dir}(t)+\lambda_{j}^{indir}(t)$. Any paper can be cited directly, but only previously cited paper can be cited indirectly. Since we focus here on the previously  uncited papers, we are interested only  in direct citations.  The model assumes that
\begin{equation}
\lambda_{j}^{dir}(t)=\eta_{j}\frac{N(t_{0}+t)R_{0}(t_{0}+t)}{N(t_{0})}A(t)
%R_{0}\tilde{A}(t).
\label{lambda}
\end{equation}
% which satisfies normalization condition $\int_{0}^{\infty}A(\tau)d\tau=1$
where    $t_{0}$ is the publication year,    $t$ is the number of years after publication,   $N(t)$ is the number of papers published in year $t$ which can potentially cite the paper $j$,   $R_{0}(t)$ is the average length of the reference lists of  these  papers,  $A(t)$ is the aging function, and $\eta_{j}$ is the papers' fitness  which is the key parameter characterizing the cited paper. The fitness  characterizes the appeal that this paper makes to the citing author after taking into account for aging and other time-dependent factors. In other words, papers's fitness captures its potential for garnering future citations. The model assumes that each paper is born with some intrinsic fitness and it does not change along the papers' lifetime.   Thus, in the context of uncited papers, our model reduces to the fitness model of Caldarelli et al. \cite{Caldarelli2002} to which we added aging and stochasticity.
%, which is the same for all papers in one discipline and published in the same year.

The probability that a paper $j$ remains uncited after $t$ years  is  $e^{-\Lambda^{dir}_{j}}$  where $\Lambda^{dir}_{j}(\eta,t)=\int_{0}^{t}\lambda^{dir}_{j}(\eta,\tau)d\tau$
%\eta_{j} R_{0}\int_{0}^{t}\tilde{A}(\tau)d\tau$
is the cumulative direct citation rate.  If  both the  number of publications and the average  reference list length  grow exponentially, namely, $R_{0}(t+t_{0})=R_{0}(t_{0})e^{\beta t}$ and $N(t+t_{0})=N(t_{0})e^{\alpha t}$, then $\frac{N(t_{0}+t)R_{0}(t_{0}+t)}{N(t_{0})}=R_{0}(t_{0})e^{(\alpha+\beta)t}$. We substitute this expression into Eq. \ref{lambda} and come to
\begin{equation}
\Lambda^{dir}(\eta, t)=\eta R_{0}(t_{0})\int_{0}^{t}A(\tau)e^{(\alpha+\beta)\tau}d\tau,
\label{Lambda}
\end{equation}
where we dropped the index $j$, for clarity.

For a  collection of papers with  different fitnesses,  all published in the same year, the fraction of uncited papers  after $t$ years (the uncitedness ratio),  is
\begin{equation}
f_{0}(t)=\int _{0}^{\infty}e^{-\Lambda^{dir}(\eta,t)}\rho(\eta)d\eta,
\label{f-0}
\end{equation}
where $\rho(\eta)$ is the  fitness distribution.% and  $\Lambda^{dir}(\eta,t)$  is given by  Eq. \ref{Lambda}.

A closely related parameter  for the same collection of papers is  $M(t)$, the cumulative mean number of citations. It  consists of the direct and indirect contributions,  $M(t)=M^{ dir}(t)+M^{indir}(t)$, the former one is directly related to the uncitedness ratio. Indeed,
\begin{equation}
M^{dir}(t)=\int _{0}^{\infty}\Lambda^{dir}(\eta,t)\rho(\eta)d\eta.
\label{mean-rate}
\end{equation}
It is important to  note that $M^{dir}(t)$ is determined by the set of  papers that can potentially cite the given collection (the whole discipline or community), while the fitness distribution $\rho(\eta)$ is determined by the collection of cited papers (discipline, journal, institution, country, etc.) for which we calculate the uncitedness ratio.  The closed expression for $M^{dir}(t)$ results after substitution of Eq. \ref{Lambda} into Eq. \ref{mean-rate},
\begin{equation}
M^{dir}(t)=\eta_{0}R_{0}(t_{0})\int_{0}^{t}A(\tau)e^{(\alpha+\beta)\tau}d\tau,
\label{Lambda1}
\end{equation}
%of direct references in a reference list of papers belonging to the discipline under consideration. The average fitness also gauges the fraction
%We found that $\eta_{0}=0.2-0.4$, depending on discipline.
where $\eta_{0}=\int _{0}^{\infty}\eta\rho(\eta)d\eta$ is the average fitness of papers in the collection for which we consider the uncitedness ratio. It should be noted that, in our formalism, the fitness always appears together with the aging function which is defined  in such a way that $\int_{0}^{\infty}A(\tau)d\tau=1$. In our previous study \cite{Golosovsky2017} we showed that, under this constraint,   $\eta_{0}$ characterizes  the average fraction of direct citations among all citations of the paper.  In what follows we introduce the reduced fitness $\tilde{\eta}=\frac{\eta}{\eta_{0}}$, in such a way that Eqs. \ref{Lambda},\ref{Lambda1} yield $\Lambda^{dir}(\eta, t)=\tilde{\eta}M^{dir}(t)$.  We substitute this expression into Eq. \ref{f-0} and find
\begin{equation}
f_{0}(t)=\int _{0}^{\infty}e^{-\tilde{\eta} M_{dir}(t)}\rho(\tilde{\eta})d\tilde{\eta}.
\label{uncited}
\end{equation}
In the particular case of  the exponential fitness distribution, $\rho(\tilde{\eta})=e^{-\tilde{\eta}}$, integration of  Eq. \ref{uncited}  yields an especially simple expression,
\begin{equation}
f_{0}(t)=\frac{1}{1+M_{dir}(t)}.
\label{exp}
\end{equation}

Equations \ref{Lambda1}, \ref{uncited} capture  the uncitedness ratio as a function of time $t$ and of the average reference list length $R_{0}$. For the limiting cases $R_{0}=0$ and $R_{0}=\infty$, Eq. \ref{uncited} yields  $f_{0}=0$ and $f_{0}=1$, correspondingly. These predictions are quite obvious since for  $R_{0}=0$  the papers do not cite one another and all of them  remain uncited, while for $R_{0}\rightarrow \infty$ the reference lists of papers are so long  that all papers will be eventually cited.

In summary, our model stipulates that the uncitedness ratio for a collection of papers  is determined by its  fitness distribution $\rho(\eta)$,   aging function $A(t)$, the average length of the reference list $R_{0}$, growth exponents $\alpha$ and $\beta$. (Besides the fitness distribution, all other parameters do not appear independently, but in a certain combination  captured by Eq. \ref{Lambda1}.) In our previous studies we studied citation dynamics of the papers belonging to  Physics, Mathematics, and Economics and measured these functions and parameters.  In what follows we check to which extent these very same  parameters and functions account for the number of papers that were not cited, in other words we verify Eqs. \ref{Lambda1}, \ref{uncited}.

\section{Measurements of uncitedness ratio for Physics, Economics, and Mathematics papers}
 We used the Clarivate WoS database, pinpointed all pure Mathematics,  all Economics papers, and  Physics papers published in 82 most important journals.  We considered only  articles, letters and notes written in English  while the overviews were excluded. We focused on the  papers published in 1984 and measured their  citation dynamics during subsequent 28 years.   Table \ref{tab:tableI} (Appendix) lists the parameters of citation dynamics for these collections.

\begin{figure}[ht]
\includegraphics*[width=0.3\textwidth]{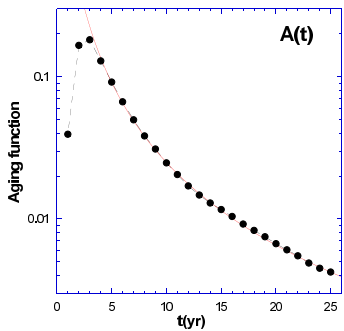}
\caption{Aging function.  The publication year corresponds to $t=1$.  The continuous red line is the empirical power-law fit, $A(t)=\frac{2.75}{(t+0.5)^{2}}$, where publication year corresponds to $t=1$.
}
\label{fig:aging-function}
\end{figure}

\begin{figure}[ht]
\includegraphics*[width=0.3\textwidth]{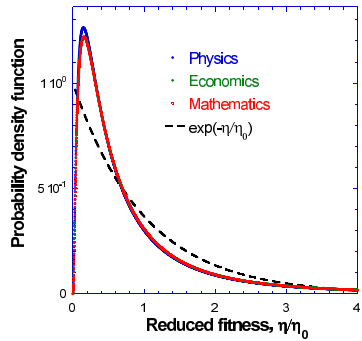}
\caption{ Reduced fitness distributions $\rho(\tilde{\eta})$ for the Physics, Mathematics, and Economics papers published in 1984  are well approximated by the log-normal distributions with very close parameters ($\sigma=1.13, 1.10,1.08$, correspondingly, and $\mu=-\frac{\sigma^2}{2}$). The dashed line shows exponential distribution, $\rho(\tilde{\eta})=e^{-\tilde{\eta}}$.
}
\label{fig:fitness}
\end{figure}

\begin{figure}[ht]
\begin{center}
\includegraphics*[width=0.3\textwidth]{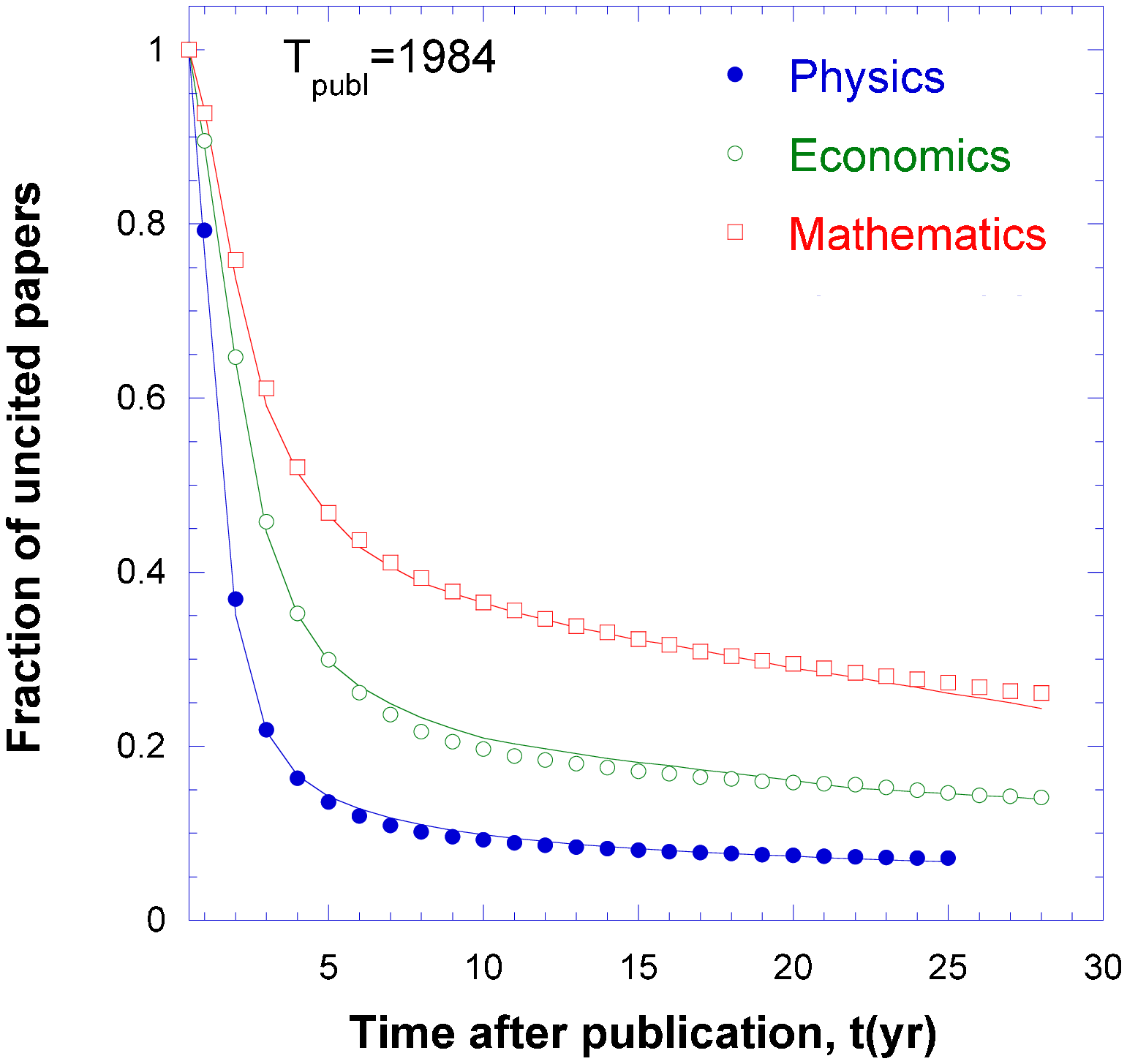}
\includegraphics*[width=0.3\textwidth]{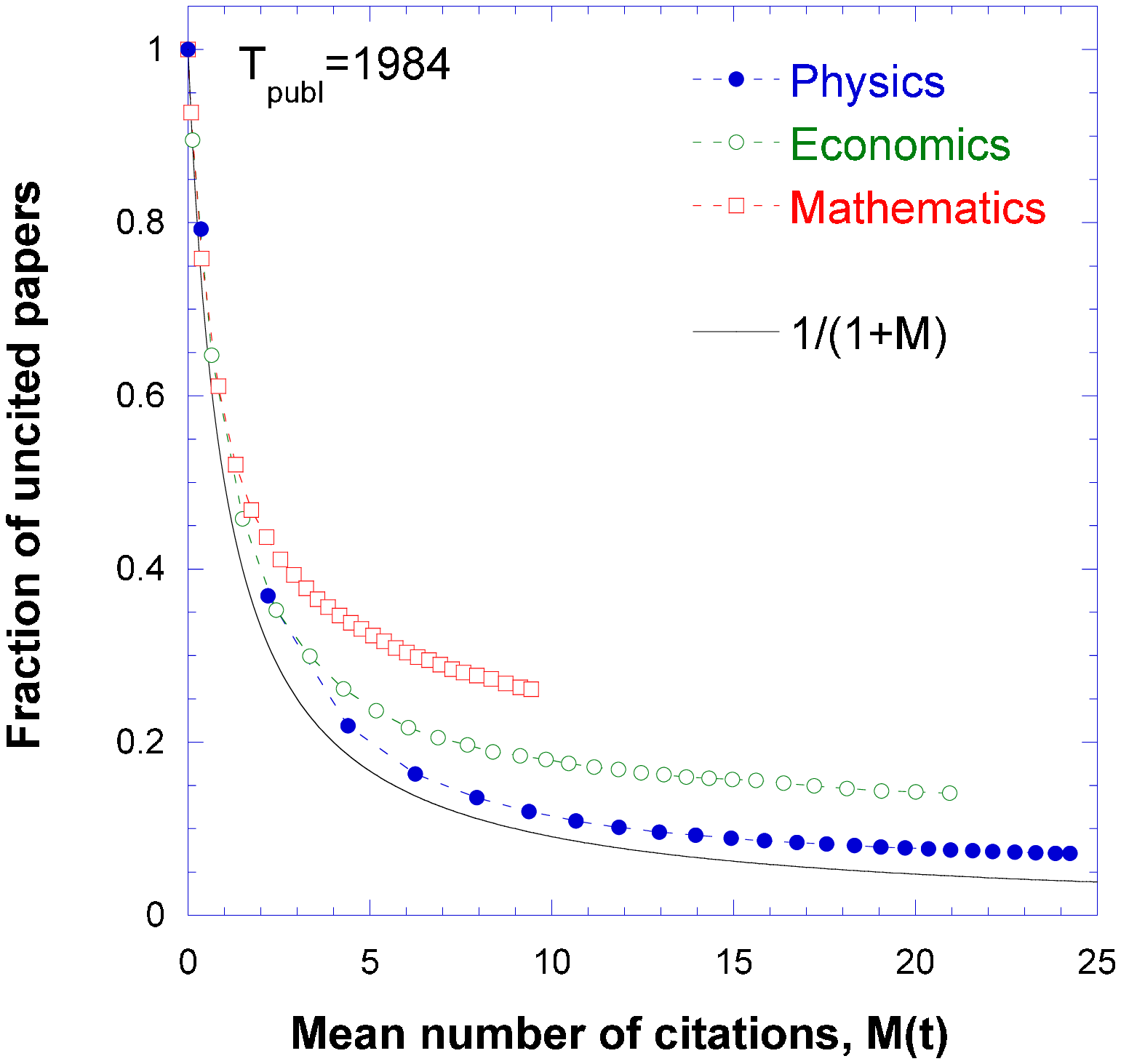}
\includegraphics*[width=0.3\textwidth]{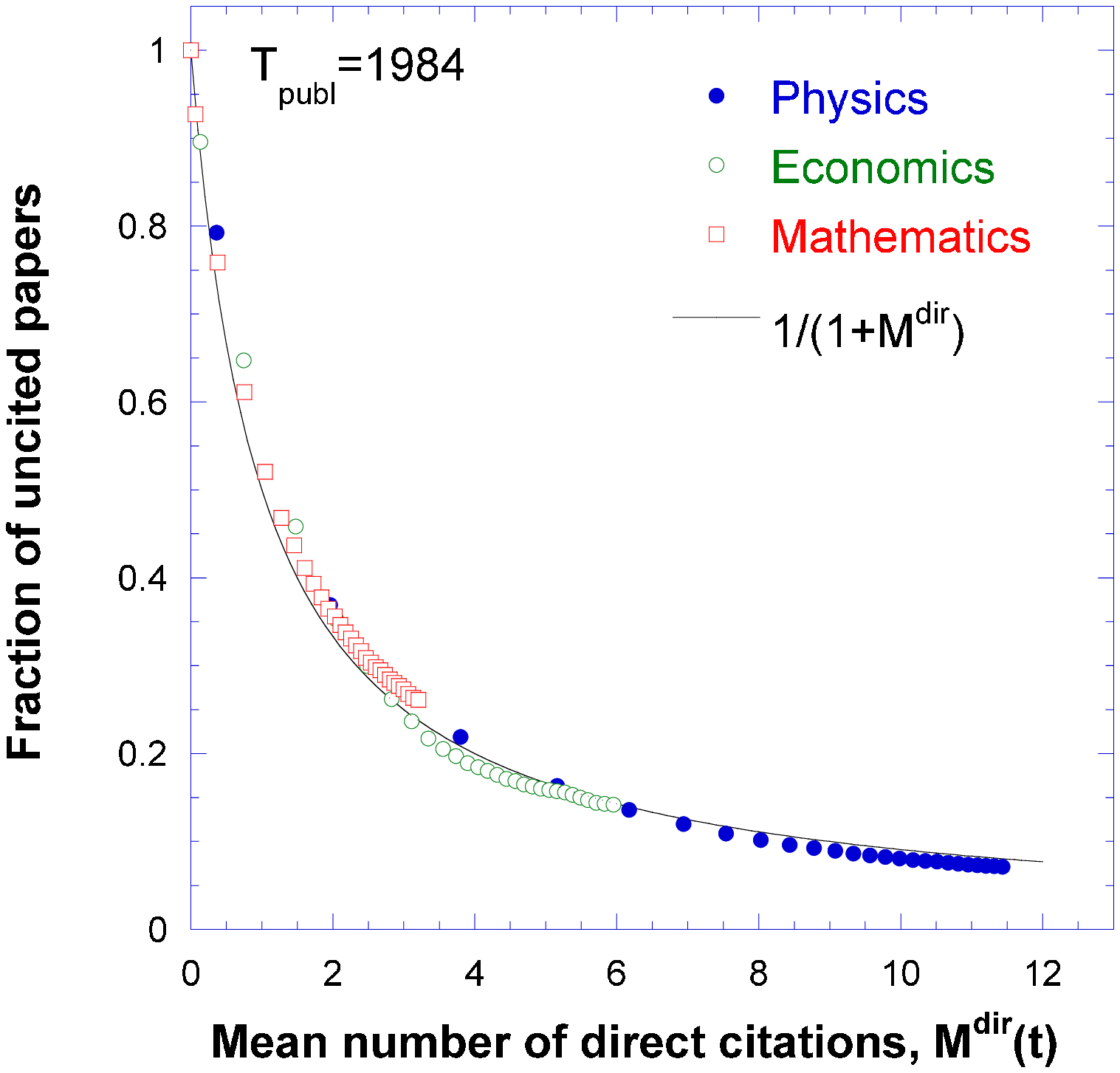}
\caption{The fraction of uncited papers, $f_{0}=\frac{N_{uncited}(t)}{N}$ where $N$ is the total number of papers belonging to some discipline and published in  year $t_{0}$, while  $N_{uncited}(t)$ is the number of uncited papers after $t$ years. (a) $f_{0}(t)$ dependences for three disciplines are markedly different. Continuous lines show results of numerical simulation based on our stochastic citation model captured by Eqs. \ref{lambda}, \ref{Lambda}. Note good correspondence to measurements.  (b) $f_{0}$ versus mean number of citations $M$.  Continuous line  shows prediction of Eq. \ref{M}, based on models of Refs. \cite{Burrell2013,Egghe2013,Hsu2012}. For $t>2$, the  data for all disciplines lie well below the model prediction. (c) $f_{0}(M_{dir})$  dependences for different disciplines collapse onto a single curve.  Here, $M_{dir}$ is the average number of direct citations given by Eqs. \ref{Lambda1}, \ref{uncited}.  Continuous line shows approximation by  Eq. \ref{exp} which  assumes exponential fitness distribution. No fitting parameters have been used.
}
\label{fig:uncited}
\end{center}
\end{figure}

Figure \ref{fig:aging-function} shows the aging function $A(t)$ which turned out to be the same for all three disciplines. It achieves maximum  after 2-3 years  and then slowly decays following the power-law dependence.
Figure \ref{fig:fitness}  shows the fitness distributions $\rho(\eta)$ found from the analysis of citation distributions. These can be well-approximated by the log-normal distributions with very similar parameters  for all three disciplines.
Figure \ref{fig:uncited}a shows the uncitedness ratio $f_{0}(t)$ for these disciplines. To compare these $f_{0}(t)$ dependences to our model,  we performed stochastic numerical simulations based on Eq. \ref{lambda} and using  measured $A(t), \rho(\eta)$  and other parameters of citation dynamics shown in Table \ref{tab:tableI}. Figure \ref{fig:uncited}a demonstrates that our simulations  capture our measurements fairly well.

Since our measurements of citation dynamics suggest that the fitness distributions for three disciplines are almost the same,  then Eq. \ref{uncited} implies that, in this case, the difference between uncitedness ratios for different disciplines is set by $M_{dir}$ alone. To demonstrate this,  we determined $M_{dir}(t)$  from Eq. \ref{Lambda1} and plotted  $f_{0}(M_{dir})$ dependences for all three disciplines together. Figure \ref{fig:uncited} shows that these dependences nearly collapse onto a single curve which is satisfactorily approximated by   Eq. \ref{exp}. Although this equation assumes an exponential fitness distribution while the actual distributions are more close to log-normal (Fig. \ref{fig:fitness}), it turns out that $f_{0}$ is not very sensitive to the shape of the fitness distribution. Thus, in the context of modeling the fraction of uncited papers,   the fitness distribution can be approximated by an exponential.   Of course, this approximation can't be used for modeling the whole citation distribution since it dramatically underestimates the  tail  which consists of high-fitness papers.

In the long time limit of 25-28 years after publication, Fig. \ref{fig:uncited}a  yields, correspondingly, 7.1$\%$,   14$\%$, and  26$\%$  uncited papers for Physics, Economics, and Mathematics papers published in 1984. However, these numbers are not final and, as   Fig. \ref{fig:uncited}b shows, the fraction of uncited papers continuously decreases and does not come to saturation. The reason for this is not only the time after publication but the slow increase in the number of publications and in the  average reference list length $R_{0}$ as well.

Previous models of uncitedness, as  summarized by Burrell \cite{Burrell2013}, assume that the author of a new paper randomly chooses  his references and makes his choice basing on some attribute of the target paper (we name it fitness). When the perspective is shifted to the cited paper, these considerations  mean that each paper has an individual citation rate. Statistical distribution of these rates has been postulated to follow  Gamma-distribution \cite{Burrell2013} or to be result of the preferential attachment rule \cite{Hsu2012}. Assuming  that  citation dynamics of each paper can be described by a  Poisson process,  the previous models \cite{Burrell2013,Egghe2013,Hsu2012} related the uncitedness ratio  for some collection of papers  to the mean number of cumulative citations for this collection, $M(t)$. In particular, for the exponential fitness distribution these models yield
\begin{equation}
f_{0}(t)=\frac{1}{1+M}.
\label{M}
\end{equation}
This prediction has a very limit range of applicability. Indeed, Fig. \ref{fig:uncited}b plots the measured uncitedness ratios, $f_{0}(t)$, versus $M(t)$.  We observe that Eq. \ref{M} accounts for these data only for a limited period of time: just a couple of years after publication. Later on, $f_{0}(t)$ dependences for  three different disciplines diverge and all lie above the curve predicted by Eq. \ref{M}. Thus, previous models fail to account for the uncitedness ratio at all times.

In contrast to previous studies \cite{Burrell2013,Egghe2013,Hsu2012}, our model assumes a much more realistic scenario of the citation process.  It takes  into account that, in filling the reference lists of their papers, the authors combine two strategies: random search (we name it direct references) and "copying" from the reference lists of the preselected papers (indirect references).  When the perspective is shifted to the cited paper, these correspond to the direct and indirect citations. While previous models related the uncitedness ratio to the average number of all citations $M(t)$, our model relates it to $M^{dir}(t)$, the average number of direct citations.
%The very fact that Eqs.  \ref{Lambda1},\ref{uncited},\ref{exp} capture the number of uncited papers is significant.  Indeed, all parameters and variables in these equations are related to citations and their dynamics, none of them is related to non-citations. Non-citations here come as a byproduct of citations and this is the Poisson statistics which makes the cited and uncited papers inseparable \cite{Burrell2013}. All this  indicates that the uncited papers constitute the inherent part of the scientific enterprize, namely, uncited papers are not unread \cite{Seglen1992,vanNoorden2017}.

\section{Conclusions}
The very fact that our model of citation dynamics captures the number of uncited papers is significant.  Indeed, all parameters and variables in this model are related to citations and their dynamics, none of them is related to non-citations. The latter come as a byproduct of citations and this is the Poisson statistics which makes the cited and uncited papers inseparable. All this  indicates that, for three disciplines we studied here, the uncited papers constitute the inherent part of the scientific enterprise, namely, uncited papers are not unread.
\appendix
\section{Parameters of citation dynamics}
To find the functions and parameters appearing in Eq. \ref{Lambda1}, we relied on the measured $M(t)$ dependence and on the reference-citation duality \cite{Golosovsky2017}.  Namely, if the  number of publications and of the reference list length both grow exponentially, then the mean number of citations is $M(t)=R(t)e^{(\alpha+\beta)t}$ where $R(t)$ is the  age distribution of references (diachronous citation distribution) and $R_{0}$ is the average reference list length at the publication year. Unlike $M(t)$, which can diverge with time, $R(t)$ converges to $R_{0}$ in the long time limit. For fitting purposes, we cast the above expression as follows
\begin{equation}
\frac{M(t)e^{-(\alpha+\beta)t}}{R_{0}}=r(t),
\label{MR}
\end{equation}
where $r(t)$ is the reduced age distribution of references which is remarkably stable over time and its variability  between different disciplines is not very pronounced \cite{Golosovsky2017,Sinatra2015,Roth2012}. For  each discipline,  we plotted the $r(t)$ dependence according to Eq. \ref{MR} with $R_{0}$ and $(\alpha+\beta)$ as fitting parameters, the criterion being convergence of $r(t)$   to unity in the long time limit. The results are shown in  Table \ref{tab:tableI}.
\begin{table}[ht]
\caption{\label{tab:tableI}%
Model parameters for the papers used in our measurements, $T_{publ}=1984$.
}
\begin{ruledtabular}
\begin{tabular}{lcccc}
Disciplines&
\textrm{Number of papers, $N$}&
\textrm{$R_{0}$\footnote{The average reference list length, as estimated from $M(t)$.}}&
\textrm{$(\alpha+\beta)$\footnote{The sum of growth exponents, as estimated from $M(t)$.}}&
\textrm{$\eta_{0}$\footnote{Average fitness, as found from the fitness distributions.}}\\
%\textrm{$R_{0}\eta_{0}$\footnote{Calculated.}}\\
%&Number of papers&Reference list length&Growth exponents&Average fitness&\\
\colrule
Physics&40195&18 & 0.045&0.49\\
Economics &3043& 8 &0.085 & 0.39\\
Mathematics &6313&3.6&0.092 & 0.435\\
\end{tabular}
\end{ruledtabular}
\end{table}
We observe that $R_{0}$  is  smaller than the actual reference list length but it should be noted  that it counts only those references that can cite the given paper and that are included in the citation database. For  Web of Science, these include only research papers and exclude books, conference proceedings, etc. The fraction of these documents in the reference lists of Physics papers is small, hence $R_{0}$ for Physics matches our independent measurements \cite{Golosovsky2017}. However,  the fraction of books and conference proceedings in the reference lists of the Economics and Mathematics papers is rather big, and that is why the effective $R_{0}$ for these disciplines is so small.  %With respect to the growth exponents, they are more or less compatible with direct measurements ($\approx 3\%$ annual growth in the reference list length and $\approx 4\%$ growth of the number of publications - \cite{Sugimoto2018}), only the exponent for Physics is too low.

 % The reference list length is probably too small but it should be noted that $R_{0}$  captures only  references to research papers  included in WoS, while the books and conference proceedings which are so abundant in Mathematics and Economics, are excluded.

 %% similar to one which  we used earlier for Physics papers published in the same year.

\bibliography{reference_master_2019_new}

\begin{thebibliography}{16}%
\makeatletter
\providecommand \@ifxundefined [1]{%
 \@ifx{#1\undefined}
}%
\providecommand \@ifnum [1]{%
 \ifnum #1\expandafter \@firstoftwo
 \else \expandafter \@secondoftwo
 \fi
}%
\providecommand \@ifx [1]{%
 \ifx #1\expandafter \@firstoftwo
 \else \expandafter \@secondoftwo
 \fi
}%
\providecommand \natexlab [1]{#1}%
\providecommand \enquote  [1]{``#1''}%
\providecommand \bibnamefont  [1]{#1}%
\providecommand \bibfnamefont [1]{#1}%
\providecommand \citenamefont [1]{#1}%
\providecommand \href@noop [0]{\@secondoftwo}%
\providecommand \href [0]{\begingroup \@sanitize@url \@href}%
\providecommand \@href[1]{\@@startlink{#1}\@@href}%
\providecommand \@@href[1]{\endgroup#1\@@endlink}%
\providecommand \@sanitize@url [0]{\catcode `\\12\catcode `\$12\catcode
  `\&12\catcode `\#12\catcode `\^12\catcode `\_12\catcode `\%12\relax}%
\providecommand \@@startlink[1]{}%
\providecommand \@@endlink[0]{}%
\providecommand \url  [0]{\begingroup\@sanitize@url \@url }%
\providecommand \@url [1]{\endgroup\@href {#1}{\urlprefix }}%
\providecommand \urlprefix  [0]{URL }%
\providecommand \Eprint [0]{\href }%
\providecommand \doibase [0]{http://dx.doi.org/}%
\providecommand \selectlanguage [0]{\@gobble}%
\providecommand \bibinfo  [0]{\@secondoftwo}%
\providecommand \bibfield  [0]{\@secondoftwo}%
\providecommand \translation [1]{[#1]}%
\providecommand \BibitemOpen [0]{}%
\providecommand \bibitemStop [0]{}%
\providecommand \bibitemNoStop [0]{.\EOS\space}%
\providecommand \EOS [0]{\spacefactor3000\relax}%
\providecommand \BibitemShut  [1]{\csname bibitem#1\endcsname}%
\let\auto@bib@innerbib\@empty
%</preamble>
\bibitem [{\citenamefont {de~Solla~Price}(1965)}]{SollaPrice1965}%
  \BibitemOpen
  \bibfield  {author} {\bibinfo {author} {\bibfnamefont {D.~J.}\ \bibnamefont
  {de~Solla~Price}},\ }\href {\doibase 10.1126/science.149.3683.510} {\bibfield
   {journal} {\bibinfo  {journal} {Science}\ }\textbf {\bibinfo {volume}
  {149}},\ \bibinfo {pages} {510} (\bibinfo {year} {1965})}\BibitemShut
  {NoStop}%
\bibitem [{\citenamefont {Seglen}(1992)}]{Seglen1992}%
  \BibitemOpen
  \bibfield  {author} {\bibinfo {author} {\bibfnamefont {P.~O.}\ \bibnamefont
  {Seglen}},\ }\href {\doibase
  10.1002/(sici)1097-4571(199210)43:9<628::aid-asi5>3.0.co;2-0} {\bibfield
  {journal} {\bibinfo  {journal} {J. Am. Soc. Inf. Sci.}\ }\textbf {\bibinfo
  {volume} {43}},\ \bibinfo {pages} {628} (\bibinfo {year} {1992})}\BibitemShut
  {NoStop}%
\bibitem [{\citenamefont {Wallace}\ \emph {et~al.}(2009)\citenamefont
  {Wallace}, \citenamefont {Lariviere},\ and\ \citenamefont
  {Gingras}}]{Wallace2009}%
  \BibitemOpen
  \bibfield  {author} {\bibinfo {author} {\bibfnamefont {M.~L.}\ \bibnamefont
  {Wallace}}, \bibinfo {author} {\bibfnamefont {V.}~\bibnamefont {Lariviere}},
  \ and\ \bibinfo {author} {\bibfnamefont {Y.}~\bibnamefont {Gingras}},\ }\href
  {http://www.sciencedirect.com/science/article/pii/S1751157709000327}
  {\bibfield  {journal} {\bibinfo  {journal} {Journal of Informetrics}\
  }\textbf {\bibinfo {volume} {3}},\ \bibinfo {pages} {296} (\bibinfo {year}
  {2009})}\BibitemShut {NoStop}%
\bibitem [{\citenamefont {Nicolaisen}\ and\ \citenamefont
  {Frandsen}(2019)}]{Nicolaisen2019}%
  \BibitemOpen
  \bibfield  {author} {\bibinfo {author} {\bibfnamefont {J.}~\bibnamefont
  {Nicolaisen}}\ and\ \bibinfo {author} {\bibfnamefont {T.~F.}\ \bibnamefont
  {Frandsen}},\ }\href {https://doi.org/10.1007/s11192-019-03064-5} {\bibfield
  {journal} {\bibinfo  {journal} {Scientometrics}\ }\textbf {\bibinfo {volume}
  {119}},\ \bibinfo {pages} {1227} (\bibinfo {year} {2019})}\BibitemShut
  {NoStop}%
\bibitem [{\citenamefont {Thelwall}(2016)}]{Thelwall2016b}%
  \BibitemOpen
  \bibfield  {author} {\bibinfo {author} {\bibfnamefont {M.}~\bibnamefont
  {Thelwall}},\ }\href
  {http://www.sciencedirect.com/science/article/pii/S1751157716300153}
  {\bibfield  {journal} {\bibinfo  {journal} {Journal of Informetrics}\
  }\textbf {\bibinfo {volume} {10}},\ \bibinfo {pages} {622} (\bibinfo {year}
  {2016})}\BibitemShut {NoStop}%
\bibitem [{\citenamefont {van Leeuwen}\ and\ \citenamefont
  {Moed}(2005)}]{Leeuwen2005}%
  \BibitemOpen
  \bibfield  {author} {\bibinfo {author} {\bibfnamefont {T.~N.}\ \bibnamefont
  {van Leeuwen}}\ and\ \bibinfo {author} {\bibfnamefont {H.~F.}\ \bibnamefont
  {Moed}},\ }\href {https://doi.org/10.1007/s11192-005-0217-z} {\bibfield
  {journal} {\bibinfo  {journal} {Scientometrics}\ }\textbf {\bibinfo {volume}
  {63}},\ \bibinfo {pages} {357} (\bibinfo {year} {2005})}\BibitemShut
  {NoStop}%
\bibitem [{\citenamefont {Noorden}(2017)}]{vanNoorden2017}%
  \BibitemOpen
  \bibfield  {author} {\bibinfo {author} {\bibfnamefont {R.~V.}\ \bibnamefont
  {Noorden}},\ }\href@noop {} {\bibfield  {journal} {\bibinfo  {journal}
  {Nature}\ }\textbf {\bibinfo {volume} {552}},\ \bibinfo {pages} {162}
  (\bibinfo {year} {2017})}\BibitemShut {NoStop}%
\bibitem [{\citenamefont {Sugimoto}\ and\ \citenamefont
  {Lariviere}(2018)}]{Sugimoto2018}%
  \BibitemOpen
  \bibfield  {author} {\bibinfo {author} {\bibfnamefont {C.~R.}\ \bibnamefont
  {Sugimoto}}\ and\ \bibinfo {author} {\bibfnamefont {V.}~\bibnamefont
  {Lariviere}},\ }\href@noop {} {\emph {\bibinfo {title} {Measuring
  Research}}}\ (\bibinfo  {publisher} {Oxford University Press},\ \bibinfo
  {year} {2018})\BibitemShut {NoStop}%
\bibitem [{\citenamefont {Hsu}\ and\ \citenamefont {Huang}(2012)}]{Hsu2012}%
  \BibitemOpen
  \bibfield  {author} {\bibinfo {author} {\bibfnamefont {J.-W.}\ \bibnamefont
  {Hsu}}\ and\ \bibinfo {author} {\bibfnamefont {D.-W.}\ \bibnamefont
  {Huang}},\ }\href {\doibase https://doi.org/10.1016/j.physa.2011.11.028}
  {\bibfield  {journal} {\bibinfo  {journal} {Physica A: Statistical Mechanics
  and its Applications}\ }\textbf {\bibinfo {volume} {391}},\ \bibinfo {pages}
  {2129 } (\bibinfo {year} {2012})}\BibitemShut {NoStop}%
\bibitem [{\citenamefont {Burrell}(2013)}]{Burrell2013}%
  \BibitemOpen
  \bibfield  {author} {\bibinfo {author} {\bibfnamefont {Q.~L.}\ \bibnamefont
  {Burrell}},\ }\href {\doibase 10.1016/j.joi.2013.03.001} {\bibfield
  {journal} {\bibinfo  {journal} {Journal of Informetrics}\ }\textbf {\bibinfo
  {volume} {7}},\ \bibinfo {pages} {676} (\bibinfo {year} {2013})}\BibitemShut
  {NoStop}%
\bibitem [{\citenamefont {Egghe}(2013)}]{Egghe2013}%
  \BibitemOpen
  \bibfield  {author} {\bibinfo {author} {\bibfnamefont {L.}~\bibnamefont
  {Egghe}},\ }\href
  {http://www.sciencedirect.com/science/article/pii/S1751157712000909}
  {\bibfield  {journal} {\bibinfo  {journal} {Journal of Informetrics}\
  }\textbf {\bibinfo {volume} {7}},\ \bibinfo {pages} {183} (\bibinfo {year}
  {2013})}\BibitemShut {NoStop}%
\bibitem [{\citenamefont {Golosovsky}\ and\ \citenamefont
  {Solomon}(2017)}]{Golosovsky2017}%
  \BibitemOpen
  \bibfield  {author} {\bibinfo {author} {\bibfnamefont {M.}~\bibnamefont
  {Golosovsky}}\ and\ \bibinfo {author} {\bibfnamefont {S.}~\bibnamefont
  {Solomon}},\ }\href {\doibase 10.1103/physreve.95.012324} {\bibfield
  {journal} {\bibinfo  {journal} {Physical Review E}\ }\textbf {\bibinfo
  {volume} {95}},\ \bibinfo {pages} {012324} (\bibinfo {year}
  {2017})}\BibitemShut {NoStop}%
\bibitem [{\citenamefont {Golosovsky}(2019)}]{Golosovsky2019}%
  \BibitemOpen
  \bibfield  {author} {\bibinfo {author} {\bibfnamefont {M.}~\bibnamefont
  {Golosovsky}},\ }\href@noop {} {\emph {\bibinfo {title} {Citation Analysis
  and Dynamics of Citation Networks}}}\ (\bibinfo  {publisher} {Springer
  International Publishing},\ \bibinfo {year} {2019})\BibitemShut {NoStop}%
\bibitem [{\citenamefont {Caldarelli}\ \emph {et~al.}(2002)\citenamefont
  {Caldarelli}, \citenamefont {Capocci}, \citenamefont {DeLosRios},\ and\
  \citenamefont {Mu{\~{n}}oz}}]{Caldarelli2002}%
  \BibitemOpen
  \bibfield  {author} {\bibinfo {author} {\bibfnamefont {G.}~\bibnamefont
  {Caldarelli}}, \bibinfo {author} {\bibfnamefont {A.}~\bibnamefont {Capocci}},
  \bibinfo {author} {\bibfnamefont {P.}~\bibnamefont {DeLosRios}}, \ and\
  \bibinfo {author} {\bibfnamefont {M.~A.}\ \bibnamefont {Mu{\~{n}}oz}},\
  }\href {\doibase 10.1103/physrevlett.89.258702} {\bibfield  {journal}
  {\bibinfo  {journal} {Phys. Rev. Lett.}\ }\textbf {\bibinfo {volume} {89}},\
  \bibinfo {pages} {258702} (\bibinfo {year} {2002})}\BibitemShut {NoStop}%
\bibitem [{\citenamefont {Sinatra}\ \emph {et~al.}(2015)\citenamefont
  {Sinatra}, \citenamefont {Deville}, \citenamefont {Szell}, \citenamefont
  {Wang},\ and\ \citenamefont {Barabasi}}]{Sinatra2015}%
  \BibitemOpen
  \bibfield  {author} {\bibinfo {author} {\bibfnamefont {R.}~\bibnamefont
  {Sinatra}}, \bibinfo {author} {\bibfnamefont {P.}~\bibnamefont {Deville}},
  \bibinfo {author} {\bibfnamefont {M.}~\bibnamefont {Szell}}, \bibinfo
  {author} {\bibfnamefont {D.}~\bibnamefont {Wang}}, \ and\ \bibinfo {author}
  {\bibfnamefont {A.-L.}\ \bibnamefont {Barabasi}},\ }\href
  {https://doi.org/10.1038/nphys3494} {\bibfield  {journal} {\bibinfo
  {journal} {Nature Physics}\ }\textbf {\bibinfo {volume} {11}},\ \bibinfo
  {pages} {791} (\bibinfo {year} {2015})}\BibitemShut {NoStop}%
\bibitem [{\citenamefont {Roth}\ \emph {et~al.}(2012)\citenamefont {Roth},
  \citenamefont {Wu},\ and\ \citenamefont {Lozano}}]{Roth2012}%
  \BibitemOpen
  \bibfield  {author} {\bibinfo {author} {\bibfnamefont {C.}~\bibnamefont
  {Roth}}, \bibinfo {author} {\bibfnamefont {J.}~\bibnamefont {Wu}}, \ and\
  \bibinfo {author} {\bibfnamefont {S.}~\bibnamefont {Lozano}},\ }\href
  {\doibase http://dx.doi.org/10.1016/j.joi.2011.08.005} {\bibfield  {journal}
  {\bibinfo  {journal} {Journal of Informetrics}\ }\textbf {\bibinfo {volume}
  {6}},\ \bibinfo {pages} {111 } (\bibinfo {year} {2012})}\BibitemShut
  {NoStop}%
\end{thebibliography}%
\end{document}